\documentclass[pra,aps,twocolumn]{revtex4}

\usepackage{bm,graphicx,amsmath,amssymb}
\usepackage{bbm}
\usepackage{placeins}
\usepackage{subfigure}

\newcommand{\abs}[1]{\left|{#1}\right|}

\newcommand{\mean}[1]{\langle#1\rangle}

\begin{document}
\title{Single-photon frequency conversion in nonlinear crystals}
\author{Susanne Blum$^1$, Georgina A. Olivares-Renter\'ia$^{2,3},$ Carlo Ottaviani$^{2,4}$, Christoph Becher$^5$, and Giovanna Morigi$^{1,2}$}
\affiliation{$^{1}$ Fachrichtung 7.1: Theoretische Physik, Universit\"{a}t des Saarlandes, D 66123 Saarbr\"{u}cken, Germany,\\
$^{2}$Departament de F\'{i}sica, Universitat Aut\`{o}noma de Barcelona, E-08193 Bellaterra, Spain,\\
$^{3}$ Center for Quantum Optics and Quantum Information, Departamento de F\'isica,
Universidad de Concepci\'on, Casilla 160-C, Concepci\'on, Chile,\\
$^4$ Department of Computer Science, University of York, YO10 5GH York, UK,\\
$^{5}$ Fachrichtung 7.2: Experimentalphysik, Universit\"{a}t des Saarlandes, D 66123 Saarbr\"{u}cken, Germany.
}

\date{\today}

\begin{abstract}
Frequency conversion of single photons in a nonlinear crystal is theoretically discussed. Losses and noise are included within a Heisenberg-Langevin formalism for the propagating photon field. We calculate the first- and second-order correlation functions of the frequency-converted light when the input is a train of single-photon pulses. This model allows one to identify the requirements on the nonlinear device so that it can be integrated in a quantum network. 
\end{abstract}

\maketitle

\section{Introduction}

The ability to process information at the single-photon level has become a crucial prerequisite for quantum communications \cite{GisinRMP,Sanguard}. Implementations of networks based on single photons face the challenge of bridging the frequency at which the nodes optimally operate with the telecom band frequency at which transmission losses in optical fibers are minimized. A possible solution is to frequency convert single-photon wave packets into and back from the telecom band wavelengths using multi-wave mixing in nonlinear devices \cite{Tanzilli2005,Ou2008,Curtz2010,Shariar:2012}. Several experiments have demonstrated frequency conversion of light, e.g., nonclassical states of light \cite{Kumar:1992} and weak coherent states, both in upconversion (infrared to visible) \cite{Albota:2004,Roussev:2004,Pelc:2011} and downconversion (visible to infrared) \cite{Curtz2010,Takesue,Pelc,Zaske2011a}. Very recently quantum frequency conversion of single photon states was demonstrated both in upconversion \cite{Tanzilli2005,Rakher2010} and downconversion \cite{Zaske2012,Ikuta2011,Ates2012}. For single photons it has been proven that some of their classical (coherence) and non-classical (photon statistics) properties are conserved in the frequency conversion process \cite{Zaske2012,Ates2012}. All of the experiments mentioned so far used three-wave mixing in nonlinear ($\chi^{(2)}$) crystals. However, frequency conversion can also be efficiently achieved making use of four-wave mixing (FWM) in $\chi^{(3)}$ materials (e.g. Bragg scattering in nonlinear fibers or FWM in atomic gases) as demonstrated for weak coherent fields \cite{Krupa:2012,Clark:2013} and heralded single photons \cite{McGuinness2010,Kuzmich2010}. We mention that the use of atomic ensembles as nonlinear media for single-photon frequency conversion is also being discussed \cite{Gogyan2008,Fleischhauer4WM}, and has been experimentally demonstrated \cite{Kuzmich2010}. This latter approach offers a smaller bandwidth  compared with nonlinear crystals, but also a better perspective for achieving coherent control \cite{Fleischhauer}.

\begin{figure}[ht!]
   \subfigure[]{
     \label{fig:Kristall_I}
 	 \includegraphics[width=5cm]{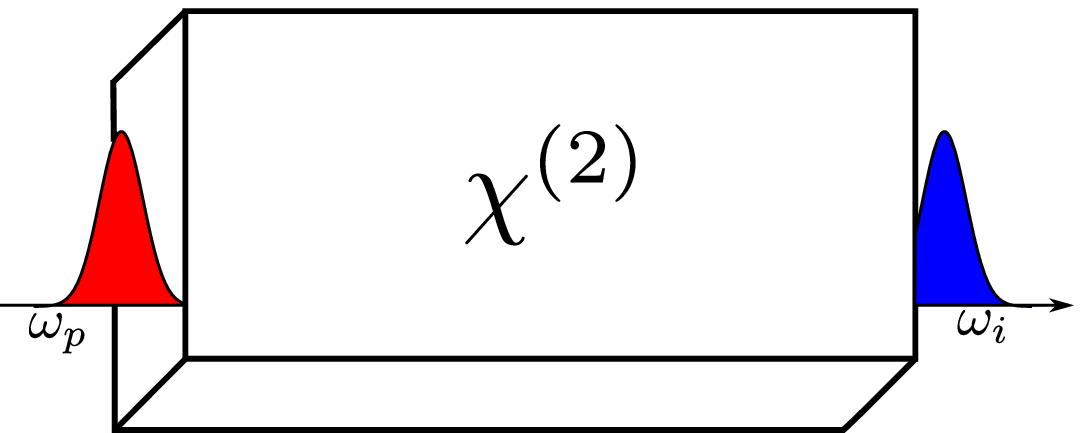}}
   \subfigure[]{
   \label{fig:Energieschema_I}
 \includegraphics[width=5cm]{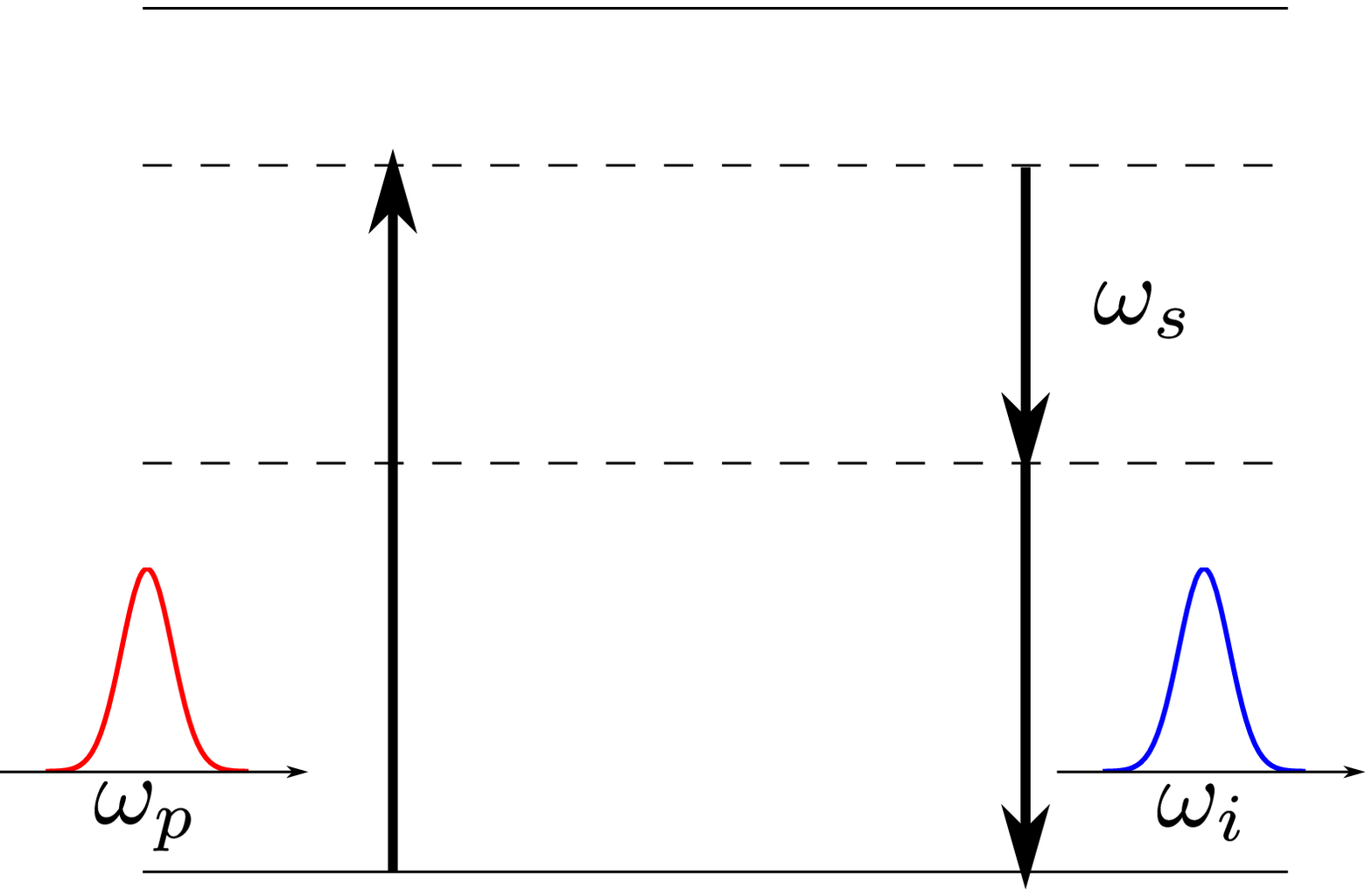}  
 }
 \caption{(color online) (a) A photon wave packet with centre frequency $\omega_p$,  typically in the visible, enters the nonlinear $\chi^{(2)}$ medium and is ideally converted into a photon wave packet with centre frequency $\omega_i$  in the low loss bands of optical fibres. (b) Sketch of the frequency conversion process. A photon at pump frequency $\omega_p$ is converted into two photons at frequency $\omega_s$ and $\omega_i$, with $\omega_p=\omega_s+\omega_i$.  The process is stimulated by a laser at frequency $\omega_s$. \label{Fig:1}} 
 \end{figure}

Since these devices shall be integrated in a quantum network, it is important to determine their efficiency at the single-photon level. This thus requires to evaluate the efficiency making use of a model describing propagation of quantum electromagnetic radiation in a medium, which we here assume to be a $\chi^{(2)}$ crystal. This model shall include losses and detrimental scattering processes, which can generate photons in the considered frequency width and/or modify the spectral and coherence properties of the frequency-converted photon at the crystal's output, and are thus a source of noise. Studies on frequency conversion between three modes have appeared for instance in Refs. \cite{Kumar:1990,Ou2008}: These works describe three-wave mixing in absence of noise and losses and do not take into account propagation inside the medium. In Ref. \cite{Christ2012} the theoretical model has been extended to a multi-mode case in which the pump field is pulsed but noise and losses are not taken into account. 

Noise and losses in the conversion process are critical in determining the efficiency of the operation at the single-photon level \cite{Pelc,Zaske2011a,Zaske2012}. Typical noise sources are optical parametric fluorescence (OPF) \cite{Pelc} and Raman scattering \cite{Zaske2012,Zaske2011a}, which can be both induced by the intense laser field required for the conversion process. To take into account realistic conditions for the frequency conversion process, in this article we discuss a theoretical model for single-photon frequency conversion in a non-linear crystal including noise and losses in the propagation dynamics. We consider a continuous pump laser and first extend the model in Refs. \cite{Kumar:1990,Ou2008} in order to include propagation of the photon field,  using the quantum theory developed in Refs. \cite{Drummond,Drummond:Raman,Hillery,Yurke,HilleryStreu}.  We identify the conditions for which a periodic exchange of single-photon excitations between the input and the target frequency band occurs. Noise and losses are then introduced within a Heisenberg-Langevin equation for the photon field propagating inside the dielectrics. 

This article is organized as follows. In Sec. \ref{Sec:Model} the theoretical model is derived. In Sec. \ref{Sec:HL} noise and losses are introduced using Heisenberg-Langevin equation. In Sec. \ref{Sec:Calculations} we apply the model to determine the first- and second-order correlation functions of the frequency transformed light when the input is a train of single photons in the optical regime. The conclusions and outlooks are presented in Sec. \ref{Sec:Conclusions}.
 
\section{Theoretical model}
\label{Sec:Model}

In this section we start from the equation for the propagation of the photon field in an ideal $\chi^{(2)}$ medium using the theoretical description developed in Refs. \cite{Drummond,Drummond:Raman,Hillery,Yurke,HilleryStreu}. Here, we identify the conditions under which a periodic exchange of photonic excitation between input and target frequency band occurs during the propagation inside the crystal. We then introduce the loss terms by means of quantum reservoirs \cite{Drummond:Raman} and solve the corresponding Heisenberg-Langevin equations of motion for the photon field. These equations allow us to determine the correlation functions of a train of photons after interaction with the crystal as a function of the input state and of the noise during the propagation dynamics.

\subsection{Hamiltonian dynamics}

The starting point of our analysis is the Hamiltonian for non-degenerate three-wave mixing in a lossless dielectric medium of length $L$, where the three coupled optical modes are at frequency $\omega_p$, $\omega_i$ and $\omega_s$. The frequencies are within a band with center frequency $\omega_p^{(0)}$, $\omega_i^{(0)}$, and $\omega_s^{(0)}$, respectively, the  corresponding wave vectors read ${\bf k_p}$, ${\bf k_i}$, and ${\bf k_s}$, such that $\omega_p=\omega_i+\omega_s$ and ${\bf k_p}={\bf k_i}+{\bf k_s}$. In the following we restrict to propagation along the positive direction of the $x$-axis, as shown in Fig. \ref{Fig:1}(a), and assume that the wave vectors are all parallel to $x$ so that we write only their modulus in the equations. Using second quantization of the electromagnetic field inside the dielectric medium according to \cite{Drummond}, we denote by $a_j$, $a_j^{\dagger}$ the annihilation and creation operators of a photon of the mode at frequency $\omega_j$ and wave vector $k_j$, with $[a_j,a_k^{\dagger}]=\delta_{jk}$. The effective Hamiltonian describing three-wave mixing can be written as
$H_0=H_{\rm emf}+H_{\rm int}$, where $H_{\rm emf}=\sum_p'\hbar\omega_p a_p^\dagger a_p+\sum_i'\hbar\omega_i a_i^\dagger a_i+\sum_s'\hbar\omega_s a_s^\dagger a_s$ contains the sum over the modes which are resonantly coupled  (denoted by $\sum'$) and 
\begin{equation}\label{HDC}
H_{\rm int}={\rm  i}\hbar{\sum_{s,p,i}}'g_{s,p,i}\mathcal F(k_p-k_i-k_s){a}_{p}^{\dag }{a}_{s}{a}_{i}+{\rm H.c. }\,
\end{equation}
describes their mutual interactions, with $\mathcal F(k)=\sin(kL/2)/(kL/2)$ (apart for a $k$-dependent phase which can be absorbed in the operators) \cite{Hillery,Christ2012}. Here, the sums over the modes are about the corresponding relevant frequency bandwidth and $g_{s,p,i}$ is the coupling parameter, which has the dimensions of a frequency and depends on the wave numbers of the modes $j=s,p,i$ \cite{Hillery,Brecht2011-1, Christ2012}. 

From Eq. \eqref{HDC} one can see that frequency-conversion of a single-photon wave packet between the frequency region centered at frequency $\omega_p^{(0)}$ and wave vector $k_p^{(0)}$, into the frequency $\omega_i^{(0)}$ and wave vector $k_i^{(0)}$ is ideally realized by a nonlinear process in which an additional photon at the difference frequency $\omega_s=\omega_p-\omega_i$ and wave vector $k_s=k_p-k_i$ is emitted. The process is sketched in Fig. \ref{Fig:1}(b). It can be enhanced by means of a laser at driving frequency $\omega_s^{(0)}$ which stimulates the photon emission process. Within this formalism this is described by assuming that the state of the mode at frequency $\omega_s^{(0)}$ is a coherent state at amplitude $\alpha_s=|\alpha_s|{\rm e}^{{\rm i}\Phi_s}$, with mean photon number $|\alpha_s|^2\gg1$. By means of a unitary transformation \cite{CT-Atom} the Hamiltonian can be written as $H^{\prime}=H_{\rm emf}+H_{\rm int}+H_F$, where 
\begin{align}\label{H:F}
H_{F}=&{\rm  i}\hbar {\sum_{p,i}}'\Omega_{p,i}^*{\rm e}^{-{\rm i}\omega_st}{a}_{i}^{\dag }{a}_{p}
+{\rm H.c. }\,,\nonumber\\
\end{align}
with $\Omega_{pi}=\Omega_{0,pi}{\rm e}^{{\rm i}\Phi_s}\mathcal F(k_p-k_i-k_s)$ and $\Omega_{0,pi}=g_{spi}|\alpha_s|$. The Hamiltonian in this form shows that this process exhibits quantum noise fluctuations due to spontaneous parametric processes, whose contribution can be made very small with respect to the stimulated emission process for laser amplitudes $|\alpha_s|^2\gg 1$. In this limit we approximate Hamiltonian $H^{\prime}$ by
\begin{equation}
\label{H:1}
H^{\prime}=H_{\rm emf}+H_F\,.
\end{equation}

\subsection{Propagation equations for the photon field}

Within the validity of this description the time evolution of the modes $p$ and $i$ is given by the Heisenberg equations governed by Eq. \eqref{H:1}, whose form can be found for instance in Ref. \cite{Christ2012}. In this article we assume that the signal is a continuous-wave laser and eliminate the explicit time dependence by performing the transformation $a_j\to a_j{\rm e}^{-{\rm i}\omega_j^{(0)}t}$, with $j=i,p$, which corresponds to moving to a rotating frame. The Heisenberg equations, in absence of noise, take the form
\begin{eqnarray}
\label{ap}
&&\dot{a}_p=-{\rm  i}\delta_p a_p-{\sum_{i}}'\Omega_{p,i}{a}_{i}\,,\\
\label{ai}
&&\dot{a}_i= -{\rm  i}\delta_i a_i+{\sum_{p}}'\Omega_{p,i}^*{a}_{p}\,,
\end{eqnarray}
with $\delta_j=\omega_j-\omega_j^{(0)}$. The case of perfect frequency conversion is found when the dependence of  the coupling rate $g_{s,p,i}$ on the frequency of the photon can be discarded, which is  verified when the input photons possess a frequency width  much smaller than the crystal's acceptance bandwidth, while the dispersion relation for the modes $p$ and $i$ is essentially linear, such that $\omega_p\simeq v_g^p k_p$ and $\omega_i\simeq v_g^i k_i$ with $k_j$ the wave vector and $v_g^j$ the group velocity. In this limit we denote by  $\psi_j(x,t)$ the field operator of a photon with spectrum in the bandwidth around $\omega_j^{(0)}$ ($j=i,p$) and $0< x<L$,
\begin{equation}
\label{phi:x}
\psi_{j}(x,t)=\frac{1}{\sqrt{L}}\sum_{j} {\rm e}^{{\rm i}(k_j-k_j^{(0)})x}a_j\,,
\end{equation}
whose dynamics are governed by the equations of motion 
\begin{eqnarray}
\label{Eq:Prop}
&&\left(\frac{\partial}{\partial t}+v_g^p \frac{\partial}{\partial x}\right)\psi_p(x,t)=-\Omega \psi_i(x,t)\,,\\
\label{Eq:Propa}
&&\left(\frac{\partial}{\partial t}+v_g^i \frac{\partial}{\partial x}\right)\psi_i(x,t)=+\Omega^*\psi_p(x,t)\,,
\end{eqnarray}
and which are valid for $0<x<L$. These equations are found by taking the time derivative of Eq. \eqref{phi:x} and using Eqs. \eqref{ai}-\eqref{ap}, after setting $\Omega\equiv\Omega_0{\rm e}^{{\rm i}\Phi_s}$ and under the assumption that the medium is sufficiently long. Periodic exchange of excitations between the $p$ and $i$ modes is realized when $v_g^p\approx v_g^i$, which is the case for the experimental conditions in Ref. \cite{Zaske2011a, Zaske2012}. In this limit the solution of Eqs. \eqref{Eq:Prop}-\eqref{Eq:Propa} read
\begin{eqnarray}
\psi_p(x,t)&=&\cos(\Omega_0\tau_x)\psi_p(0,t-\tau_x)\\
& &-e^{i\Phi_s}\sin(\Omega_0\tau_x)\psi_i(0,t-\tau_x)\,,\nonumber\\
\psi_i(x,t)&=&\cos(\Omega_0\tau_x)\psi_i(0,t-\tau_x)\\
& &+e^{-i\Phi_s}\sin(\Omega_0\tau_x)\psi_p(0,t-\tau_x)\,,\nonumber
\end{eqnarray}
where $\tau_x=x/v_g$. The equations are written as a function of the field at $x=0$ and are valid for times $t\ge \tau_x$. They describe periodic conversion of a $p$-photon into a $i$-photon and vice versa. The period is fixed by the frequency $\Omega_0$, and thus can be controlled by the intensity of the driving laser. Therefore, by setting the laser intensity $\Omega_0=(2n+1)\pi v_g/(2L)$ (with $n \in \mathbb{N}$), one has perfect conversion at the crystal end $x=L$. 

Equations~\eqref{Eq:Prop} and~\eqref{Eq:Propa} are valid under a series of conditions. In first place, we have neglected noise effects, which will be treated in the next section. We also neglected dispersion during the propagation inside the nonlinear crystal, which relies on the fulfillment of the inequality
\begin{equation}
\frac{1}{2}\frac{\partial^2\omega_{p,i}}{\partial k^2}\Bigl|_{k=k_{p,i}}\Delta_k^2T_{\rm int}\ll 1\,,
\end{equation}
with $T_{\rm int}$ the interaction time, $T_{\rm int}\sim L/v_g$, and $\Delta_k$ is the typical dispersion around the wave vector mean value. We also assumed that the group velocities of idler and pump are equal, which practically implies that 
\begin{equation}
|v_g^p-v_g^i|\Delta_kT_{\rm int}\ll 1\,.
\end{equation}
When the bandwidth of the input photon is sufficiently small, then $\Delta_k$ is determined by the interaction, $\Delta_k\approx \Omega_0/v_g$. Using $\Omega_0L/v_g=(2n+1)\pi/2$, the two inequalities to fulfill become
\begin{eqnarray}
&&(2n+1)\frac{\pi^2}{8}\frac{1}{v_gL}\frac{\partial^2\omega_{p,i}}{\partial k^2}\Bigl|_{k=k_{p,i}}\ll 1,\\
&&\frac{(2n+1)\pi}{2}\frac{|v_g^p-v_g^i|}{v_g}\ll 1.
\end{eqnarray}
The derivation of Eqs. \eqref{Eq:Prop} and~\eqref{Eq:Propa} also relies on the assumption that the length of the medium $L$ is much larger than the photon coherence length, $L\Delta_k\gg 1$. This thus leads to the choice of numbers $n\gg 1$, thus setting a lower bound to the above inequalities. Another approximation we made was to neglect spontaneous down-conversion processes which correspond to the absorption of a pump photon followed by the spontaneous emission of a signal and a idler photon. These processes are also detrimental for frequency down-conversion purposes, since the idler photon has a larger line width due to correlations with the signal photon. They can be neglected provided that the number of spontaneous emission events during propagation of a single photon are negligible, which is verified when the corresponding rate $\gamma$ fulfills the inequality $\gamma L/v_g\ll 1$.

Before concluding this section, we provide the commutation relations of the field operators. At equal times, $[\psi_j(x,t),\psi_\ell^{\dagger}(x',t)]=\delta_{j\ell}\tilde{\delta}(x-x')$, where $\tilde{\delta}(x)\approx \sin(\Delta k_j x/2)/(\pi x)$ is a smoothed version of a Dirac-delta function due to the band-limited Fourier transform: Its width scales with $1/\Delta k_j$, and $\Delta k_j$ is the half-width of the distribution of modes with wave numbers about the mode at $k_j^{(0)}$. We shall assume that this smearing effect is not pronounced (which is correct for $n\gg 1$) so that we can treat $\tilde{\delta}(x)$ as a Dirac-delta function~\cite{Hillery}. The commutation relation at different times takes a simple form when the dispersion relation is linear, and reads $[\psi_j(x,t),\psi_\ell^{\dagger}(x',t')]=\delta_{j\ell}\tilde{\delta}(x-x'-v_g(t-t'))$.

\section{Heisenberg-Langevin equations}
\label{Sec:HL}

Noise and losses are introduced in the propagation equations by means of loss and Langevin terms, which are derived after assuming that the photon fields couple with quantum reservoirs \cite{Drummond:Raman}. These processes are for simplicity here assumed to be Markovian, such that the propagation equation in presence of noise reads
\begin{eqnarray}
&\left(\frac{\partial}{\partial t}+v_g^p \frac{\partial}{\partial x}\right)\psi_p(x,t)=-\Omega\psi_i(x,t)-\kappa_p\psi_p(x,t)+L_p(x,t)\,,\nonumber\\
&\label{Eq:Prop:1}\\
&\left(\frac{\partial}{\partial t}+v_g^i \frac{\partial}{\partial x}\right)\psi_i(x,t)=\Omega^*\psi_p(x,t)-\kappa_i\psi_i(x,t)+L_i(x,t)\,.\nonumber\\
&\label{Eq:Prop:2}
\end{eqnarray}
Here, $\kappa_{j=i,p}$ is the difference between the total loss and gain rates, and can thus be either positive or negative, while $L_{j=i,p}(x,t)$ are sums of Langevin operators for different kind of noise which are uncorrelated from one another \cite{CT-Atom} and will be specified below. The solutions of Eqs. \eqref{Eq:Prop:1}-\eqref{Eq:Prop:2} read
\begin{align}
&\psi_p(x,t)=
f_{1p}(\tau_x)\psi_p(0,t-\tau_x)-f_{2}(\tau_x)\psi_i(0,t-\tau_x)\nonumber\\
&+\int_0^{\tau_x} \left[f_{1p}(\tau_1)L_p(x_1,t-\tau_1)-f_2(\tau_1)L_i(x_1,t-\tau_1)\right] {\rm d}\tau_1\,,\\
&\psi_i(x,t)=f_{1i}(\tau_x)\psi_i(0,t-\tau_x)+ f_2^*(\tau_x) \psi_p(0,t-\tau_x)\nonumber\\
&+\int_0^{\tau_x} \left[f_{1i}(\tau_1)L_i(x_1,t-\tau_1)+ 
f_2^*(\tau_1)L_p(x_,t-\tau_1)\right] {\rm d}\tau_1\,,\label{psi:i:noise}
\end{align}
where $x_1=x_1(\tau_1)>0$, 
$$x_1(\tau_1)=x-v_g\tau_1\,,$$ with $\tau_1\le \tau_x$ and $\tau_x=x/v_g\le t$. The time-dependent coefficients take the form 
\begin{align*}
&f_{1p}(y)=e^{-\kappa_s y}\left(\cos\left(\theta y\right) - \frac{\kappa_D}{\theta}\sin\left(\theta y\right)\right)\,,\\
&f_{1i}(y)=e^{-\kappa_s y}\left(\cos\left(\theta y\right) + \frac{\kappa_D}{\theta}\sin\left(\theta y\right)\right)\,,\\
&f_2(y)=\frac{\Omega}{\theta}e^{-\kappa_s y}\sin\left(\theta y\right)\,,
\end{align*}
with $\theta=(\abs{\Omega}^2-\kappa_D^2)^{1/2}$, $\kappa_D =(\kappa_p-\kappa_i)/2$, and $\kappa_s =(\kappa_i+\kappa_p)/2$.

Let us now discuss the noise sources which are accounted for in the following. These are typically (i) generic losses, where the $p$- or $i$-photon can simply be absorbed by the medium, (ii) higher nonlinear optical processes, such as Optical Parametric Fluorescence (OPF), where a $s$-photon can generate a $i$-photon and a photon at another wave length, and (iii) Raman scattering, in which $s$-photons are converted into $i$-photons by exchanging phonons with the bulk. While it is plausible that the first two types of noise can be described by means of Markovian processes, Raman scattering is instead known to be non-Markovian. This leads to integro-differential Heisenberg-Langevin equations, whose solution requires the knowledge of the corresponding noise spectrum~\cite{Drummond:Raman}. The following treatment restricts for convenience to Markovian noise. This provides a first handable solution for estimating the quantities of interest as a function of the noise threshold (it can be extended to include a full description of Raman noise, following the lines of Ref. \cite{Drummond:Raman}). For these assumptions the rates $$\kappa_j=\sum_{r_j}\left(\kappa_{r_j}^{(+)}-\kappa_{r_j}^{(-)}\right)$$ are the sum of the individual loss ($+$) and gain ($-$) rates, where the label $r_j$ refers to the specific process. The corresponding Langevin forces are $L_j(x,t)=\sum_{r_j}L_{r_j}(x,t)$, have mean value $\langle L_{r_j}(x,t)\rangle=0$ and the two-point correlations read
\begin{align}
\label{HL:2point}
&\langle  L_{r_j'}^{\dagger}(x',t')L_{r_j}(x,t)\rangle=\kappa_{r_j}^{(-)}\delta_{r_j,r_j'}\delta(t-t')\delta(x-x')\, ,\\
&\langle  L_{r_j'}(x',t')L_{r_j}^{\dagger}(x,t)\rangle=\kappa_{r_j}^{(+)}\delta_{r_j,r_j'}\delta(t-t')\delta(x-x')\,,
\label{HL:2point:1}
\end{align} 
where the average is taken over the state of the reservoirs, which is assumed to be unaffected by the coupling with the electromagnetic field within the dielectric medium, as is consistent with the Markov approximation. A microscopic theory for the coupling between the system and the quantum reservoirs allows one to establish a relation between the rates $\kappa_{r_j}$ and the physical parameters determining the process \cite{CT-Atom}. 
 
\section{Coherence properties of the frequency converted photon}
\label{Sec:Calculations}

The formalism just derived allows one to analytically determine the state of the outcoming photon, therefore to theoretically predict the results of homodyne tomo\-graphy \cite{Opatrny}. The level of noise, however, must be calibrated with the noise measured in the experiment. For this purpose we evaluate the first- and second-order correlation functions of a train of single photons which is injected into the $\chi^{(2)}$ medium in presence of typical noise sources  \cite{Pelc,Zaske2012, Zaske2011a}. We first define the input state: When reflection at the boundary of the dielectric medium can be neglected and the photon spectral properties are within the transmission band of the medium, the field at $x=0$ is proportional to the external incident photon, and the photon at the output of the dielectrics is proportional to the propagated solution till $x=L$ \cite{Yurke,HilleryStreu}. The first- and second-order correlation functions are thus evaluated at $x=L$ within the dielectric medium, and their values are plotted using parameters which are compatible with existing experiments \cite{Pelc,Zaske2012, Zaske2011a}.

\subsection{Input state}

Be the input field a train of single-photon Gaussian pulses with width in $k$-space $\Delta_k$ (hence with coherence length $\Delta_x=1/\Delta_k$), which propagates from left to right outside the dielectrics (at $x<0$). We fix the repetition rate $r=1/T_{\rm rep}$ so that the photons are spatially well separated, $v_gT_{\rm rep}\Delta_k=v_gT/\Delta_x\gg 1$, so that the (non-normalized) input state for $t<0$ reads $|\Phi_{\rm in}\rangle_t=\otimes_{j=1}^N\Phi_j^\dagger(t)|vac\rangle_{x<0}$, where $|vac\rangle_{x<0}$ is the vacuum state for the electromagnetic field in the volume at $x<0$ and 
\begin{equation}
\label{input}
\Phi_j^\dagger(t)=\int_{-\infty}^0{\rm d}x \,\frac{{\rm e}^{{\rm i}k_p^{(0)}(x-jx_0-ct)}}{(\sqrt{\pi}\Delta_x)^{1/2}}\,{\rm e}^{-\frac{(x-jx_0-ct)^2}{2\Delta_x^2}}\psi_p^\dagger(x)\,,
\end{equation}
with $x_0=-cT_{\rm rep}<0$ and $\psi_p(x)$ is the photon field at $x<0$ for frequencies within the $p$-band.  In this limit operators corresponding to different photon wave packets fulfill the commutation relation $[\Phi_j(t),\Phi_k^{\dagger}(t)]=\exp(-(j-k)^2c^2T_{\rm rep}^2/\Delta_k^2)\approx \delta_{jk}$ for $t\le 0$. We further assume that the photon spectral properties are within the transmission band and that the transmission coefficient is uniform, so that the field inside the dielectric medium can be assumed to be proportional to the input photon (Note that these assumptions can be lifted, but then one shall resort to a description in $k$- rather than $x$-representation, as in Ref. \cite{HilleryStreu}). The input field is the photon field at the boundary $x=0^-$, assuming unit transmission.

\subsection{First- and second-order correlation functions of the frequency-converted field}

We now provide some quantities of the frequency converted field at the other end of the crystal, $x=L$. We here analyze the mean photon number of the field $i$,  the first- and second-order correlation functions. 

The mean number of photons at $x=L$ and time $t$
\begin{align}
n_{i}(L,t)&=\mean{\psi_{i}^{\dag}(L,t)\psi_{i}(L,t)}\,,
\label{eq:photzahlg}
\end{align}
is here evaluated over the input state \eqref{input}. For the dynamics given by Eq. \eqref{Eq:Prop} it reads 
\begin{align}
n_{i}^{\rm ideal}(L,t)=\sin^2(\Omega_0\tau)n_p(0,t-\tau)\,,
\label{eq:photzahl}
\end{align}
where $\tau=L/v_g$ and $n_p(0,t)=\langle \psi_p^\dagger(0,t)\psi_p(0,t)\rangle$ is the mean photon number per time in the $p$-band at the input. In presence of noise, we use operators \eqref{psi:i:noise} in Eq. \eqref{eq:photzahlg} and obtain the expression
 \begin{align}
 \label{ni}
n_i(L,t)=\abs{f_2(\tau)}^2 n_p(0,t-\tau)+I(L;t,t)\,,
\end{align}
where the second term reads 
\begin{align}
&I(x;t,t+\Delta t)=\int_0^{\tau_x}{\rm d}\tau_1\int_0^{\tau_x}{\rm d}\tau_2\delta(\Delta t-(\tau_2-\tau_1))\delta(x_2-x_1) \nonumber \\
&\times\left(f_{1i}^*(\tau_1)f_{1i}(\tau_2)\sum_{r_i}\kappa_{r_i}^{(-)}+f_{2}(\tau_1)f_{2}^*(\tau_2)\sum_{r_p}\kappa_{r_p}^{(-)}\right)
 \,,
\end{align}
where we have used Eqs. \eqref{HL:2point}-\eqref{HL:2point:1}. The contribution of this latter term becomes relevant in presence of gain. When the medium is only lossy, instead, this term vanishes and the mean number of photon decreases with the factor $p_i=|f_2(\tau)|^2$. This factor gives the probability that a photon is generated by frequency conversion in the $i$-band, which is maximal for $\theta\tau={\rm atan}(\theta/\kappa_s)$. This equation reduces to $\theta\tau=\pi/2$ when $\kappa_2\ll \Omega$. In this limit $p_{i,m}=(\Omega_0/\theta)^2\exp(-\pi \kappa_s/\theta)$.  \\

The first-order correlation function is given by the expression
\begin{align}
g^{(1)}(L,t,\Delta t)={\rm Re}\{\mean{\psi^{\dag}_i(L,t)\psi_i(L,t+\Delta t)}\}\,, \label{g1function}
\end{align}
and is here reported in non-normalized form. In the ideal case it reads $g^{(1){\rm ideal}}(L,t,\Delta t)=\sin^2(\Omega_0L/v_g)g^{(1)}_p(0,t-\tau,\Delta t)$, where $g^{(1)}_p(0,t,\Delta t)={\rm Re}\{\mean{\psi^{\dag}_p(0,t)\psi_p(0,t+\Delta t)}$ is the first-order correlation function of the input field. In presence of noise, it takes the form
\begin{align}
\label{Eq:g:1}
g^{(1)}(L,t,\Delta t)&=g^{(1)}_0(L,t,\Delta t)+{\rm Re}\{I(L;t,t+\Delta t)\}\,,
\end{align}
with
\begin{equation}
\label{g1:scale}
g^{(1)}_0(L,t,\Delta t)=p_i g^{(1)}_p(0,t-\tau,\Delta t)\,.
\end{equation}
In particular, when the noise is due to losses, the first-order correlation function is the same as the one of the input photon (apart for the factor $p_i$ multiplying the whole expression). Its form can change when $I(L;t,t+\Delta t)$ is different from zero, as it additionally contains the contribution of the photons generated by noise processes. For the Markovian  noise we choose here, $I(L;t,t+\Delta t)$ vanishes for $|\Delta t|>0$.

The non-normalized second-order correlation function is 
\begin{align}\label{g2funktion}
 g^{(2)}(L,t,\Delta t)=\mean{\psi_i^{\dag}(L,t)\psi^{\dag}_i(L,t+\Delta t)\psi_i(L,t+\Delta t)\psi_i(L,t)}\,,
\end{align}
and takes the form
\begin{align}\label{g2funktion}
&g^{(2)}(L,t,\Delta t)=g^{(2)}_0(L,t,\Delta t)\nonumber\\
&+2p_i{\rm Re}\{\mean{\psi^{\dag}_p(0,t)\psi_p(0,t+\Delta t)}I(L;t+\Delta t,t)\}\nonumber\\
&+p_i(n_p(0,t-\tau)+n_p(0,t+\Delta t-\tau))I(L;t,t)\nonumber\\
&+I(L;t,t+\Delta t)I(L;t+\Delta t,t) +I(L;t,t)^2\,,\nonumber 
\end{align}
where
\begin{equation}
\label{g2:scale}
g^{(2)}_0(L,t,\Delta t)=p_i^2g^{(2)}_p(0,t-\tau,\Delta t)\,
\end{equation}
and $g^{(2)}_p(0,t,\Delta t)=\mean{\psi_p^{\dag}(0,t)\psi^{\dag}_p(0,t+\Delta t)\psi_p(0,t+\Delta t)\psi_p(0,t)}$. In Eq. \eqref{g2funktion} we have used that the noise is stationary. When the medium is only lossy, then the second-order correlation function is the same as the one of the input photon, except for the overall factor $p_i^2$, which is the probability that two photons at the input are frequency-converted into two photons in the i-band. Noise such $I\neq 0$, like OPF and Raman processes, gives rise to additional terms. One of the overall effect is to reduce the antibunching, as clearly visible after setting $\Delta t=0$ in Eq. \eqref{g2funktion}. For $\Delta t=T_{\rm rep}>\tau$, in particular, the second term on the right-hand side of Eq. \eqref{g2funktion} vanishes and antibunching is reduced by processes arising from the coincidence measurements between frequency-converted photons and noise photons, as well as between noise photons. 

\begin{figure}[ht!]
\includegraphics[width=8cm]{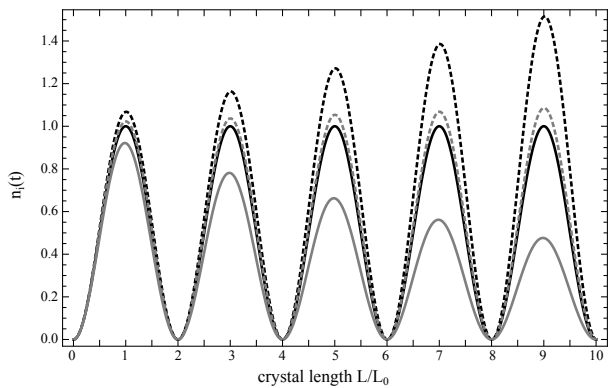}
\caption{Photon rate $n_i(t)$, in units of ${\rm max}(n_i^{\rm ideal}(t))$, versus crystal length $L$ evaluated from Eq. \eqref{ni} for the input state in Eq. \eqref{input} taking Gaussian wave packets of temporal width 1~nsec and repetition rate $10^7{\rm sec}^{-1}$. The crystal length is given in units of the minimal length $L_0=(\pi/2)v_g/\Omega_0$, for which perfect conversion is expected. The black solid line corresponds to the photon rate in the ideal case, Eq. \eqref{eq:photzahl}; the grey solid line is found when $\kappa_i=\kappa_p=0.03\cdot\Omega_0$, corresponding to 8\% losses; the gray (black) dotted line is found for 8\% losses and $\kappa_{\rm gain}=0.06\cdot \Omega_0$, corresponding to SNR 100:1 (respectively: $\kappa_{\rm gain}=0.08\cdot \Omega_0$, such that SNR is 20:1). 
}
\label{Fig:n}
\end{figure}

In order to illustrate these predictions, we apply them to experimentally relevant situations. We take photonic wave packets with temporal width of 1~nsec and repetition rate $T_{\rm rep}=10^7{\rm sec}^{-1}$, consistent with \cite{Zaske2011a,Zaske2012}. In order to minimize detrimental effects, one would try to minimize the length of the crystal. For a crystal length of the order of $L\sim 5.3\,$cm, we choose $\Omega_0 = 2\pi \times 0.6\,{\rm GHz}$  with $\Omega_0 L/v_g=\pi/2$, where we assumed $v_g\approx c/2.2$. Figure~\ref{Fig:n} displays $n_i(t)$, Eq. \eqref{ni}, as a function of the crystal length $L$. The curves compare the ideal case, Eq. \eqref{eq:photzahl}, with the case in which 8\% losses are present, and then when also gain processes (due for instance to Raman scattering or optical parametric fluorescence) are present. For these gain processes we take values of the signal-to-noise (SNR) ratios which are 20:1 \cite{Zaske2012} and 100:1 \cite{ZaskePhD}, respectively. The observed increase of the maximum as a function of the propagation lengths is due to noise photons, which are generated during the propagation time. 

\begin{figure}[ht!]
\includegraphics[width=8cm]{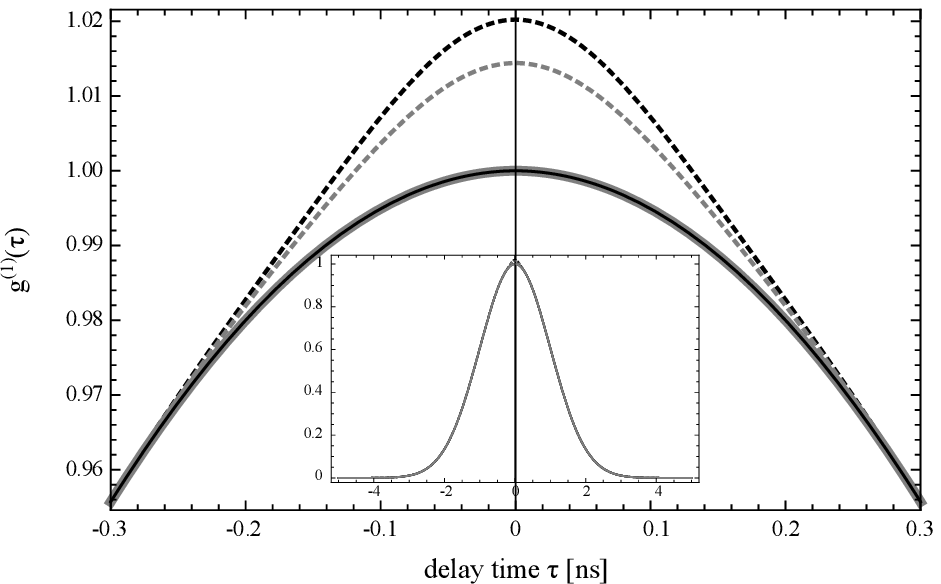}
\includegraphics[width=8cm]{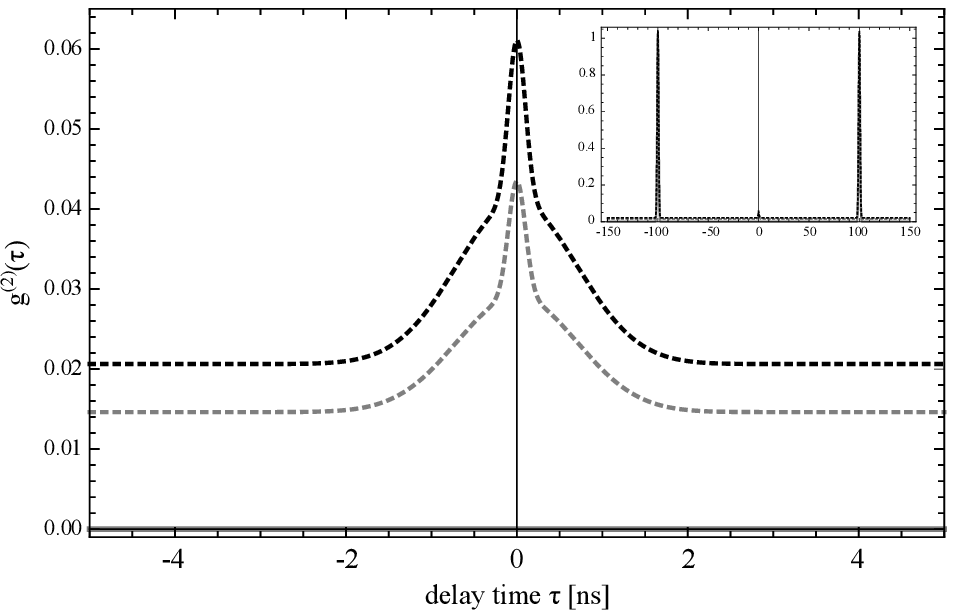}
\caption{(a) First- and (b) second-order correlation function. The first-order correlation functions is evaluated using Eq. \eqref{Eq:g:1}, rescaled by the maximum of $g^{(1)}_0$, Eq. \eqref{g1:scale}. The second-order correlation function is calculated from Eq. \eqref{g2funktion}, rescaled by the maximum of $g^{(2)}_0$, Eq. \eqref{g2:scale}. The parameters and line styles are the same as in Fig. \ref{Fig:n}, where we choose $\Omega_0=2\pi \times 0.6\,{\rm GHz}$. The width of the noise photons is here assumed to be of the order of 50~GHz, which is consistent with the measurement in \cite{Zaske2012,ZaskePhD}. }
\label{Fig:g1}
\end{figure}

Figures~\ref{Fig:g1}(a) and (b) display the first- and second-order correlation function as a function of $\Delta t$ for the input state in Eq. \eqref{input}. The different curves correspond to different noise sources and strength. The first-order correlation function has been rescaled by the probability $p_i$, and is in general is unaffected by losses. Gain gives rise to a contribution about $\Delta t=0$, which corresponds to a peak with the temporal width of an incoherently generated photon. Inspection in the behaviour of the second-order correlation function shows how antibunching degrades due to the effect of gain processes, giving rise to a background noise and a peak at zero-time delay due to the generation of incoherent photons. 

\section{Concluding remarks}
\label{Sec:Conclusions}

We have presented a model which describes frequency conversion of a photon field propagating inside a $\chi^{(2)}$-medium, when the medium is continuously pumped by a laser. We have  identified the conditions for which ideal frequency conversion is achieved. Noise and losses have been introduced within a Heisenberg-Langevin equations formalism. For simplicity, we have assumed the noise to be Markovian. The model here derived can be applied to calculate all coherence properties of the generated photon, including for modeling quantum correlations revealed by  Franson interferometry  \cite{Franson,Tanzilli2005}. Here, we have used it to determine the first- and second-order correlation functions of the frequency-converted field, which are currently measured in experiments dealing with single-photon frequency conversion. 

This theoretical description can be extended to to take into account specific properties once the spectral features of the noise of the medium is known, as done for instance in \cite{Drummond:Raman}. In this case the equations should be conveniently cast in $k$-representation. One could also consider to minimize detrimental effects by using a pulsed drive, in the spirit of the work in Ref. \cite{Christ2012}. Here,  the frequency spectrum of the pulses could be so shaped in order to optimize the frequency-conversion process for a given type of input photon. These ideas can be integrated with optimal control techniques \cite{OCT}: Such approach could possibly reduce the interaction time, and thereby detrimental effects which are unavoidable in bulk materials.  

\acknowledgements
The authors acknowledge discussions with Andreas Christ, Andreas Lehnard, Christine Silberhorn, and Sebastian Zaske. This work has been supported by the BMBF (QuORep, Contract Nos. 01BQ1011 and 16BQ1011), the Spanish Ministry of Research and Science, and by the German Research Foundation (DFG).

\end{document}